\documentclass[aps,prd,preprint,superscriptaddress,showpacs,showkeys]{revtex4}
\usepackage{amssymb,amsmath}
\usepackage{graphicx} 
\usepackage{dcolumn}  
\usepackage{epsfig}   
\usepackage{pstricks}
\usepackage{ifpdf}
\begin{document}


\title{Why does Schr\"odinger's cat refuse to be quantic}


%
%

\author{D-M. Cabaret}
\affiliation{Ecole biblique et arch\'eologique francaise de Jerusalem}
\email[]{dominiquemarie.cabaret@gmail.com}
\author{T. Grandou}
\affiliation{Universit\'{e} de Cote d'Azur,\\ Institut de Physique de Nice, UMR CNRS 7010, 1361 routes des Lucioles, 06560 Valbonne, France}
\email[]{thierry.grandou@inphyni.cnrs.fr}
\author{E. Perrier}
\affiliation{Universit\'e catholique de Toulouse\\ 
1, Impasse Lacordaire\\ 
Toulouse\\ 
31400 France}
\email[]{perrier@revuethomiste.fr}


\date{\today}

\begin{abstract} 
Both at formal and philosophical levels, decades of efforts have been devoted to a deciphering of the `quantum enigma', the `crazy way quantum objects behave', to quote R.P. Feynman's words. We posit that the enigma cannot recede without resorting to a thorough questioning of the quantum ontology. In the current paper this questioning is introduced through a revisitation of the famous Schr\"odinger's cat paradox, and of the main attempts at solving it. Our proposal thereof, is that the quantum world enjoys a sound physical reality (complying, in particular, with the scientific requirement of experimental reproducibility), but also, that this physical reality calls for an enlarged, and however precise notion of `what is real', such that what is real and actual can be distinguished from what is real but potential. 
\end{abstract}


\maketitle

\section{\label{SEC:1}Introduction}

Since its discovery and completion around 1925, Quantum Mechanics ($QM$) has never ceased to lend itself to endless debates concerning its interpretation, whereas, on the contrary, its efficiency suffers no questioning. The set of $QM$ theoretical predictions is so supported by generations of experimental tests that no doubt exists today concerning the extraordinary relevance of $QM$ to the description of the microphysical world.



\par
At a formal level, it is remarkable that $QM$ amounts to just a few axioms which, if not complying with ordinary intuition, are simple enough to be the matter of an undergraduate course. Moreover, contrary to what was first thought, that $QM$ laws would apply only in a statistical sense to a large number of {{identically prepared systems}}, it is known at present that elementary quantum systems are also perfectly described by the laws of Quantum Mechanics. This state of affairs can even be considered as a kind of second quantum revolution, initiated in the last quarter of the XXth century \cite{Gisin1}. Concerning these $QM$ statistical aspects, deep theorems obtained at a later time in 1967, 2006 and 2009, themselves initiated by the famous {Gleason Theorem} \cite{Gleason}, have elucidated a $QM$ peculiarity which definitely lacks any classical equivalent \cite{KSKC}. In effect, the statistical character of a quantum system is \emph{inherent} and manifest in the case of a single and simple quantum entity, in a radical contradistinction to the classical situation where it always reflects a lack of knowledge on a system endowed with a large number of degrees of freedom.

\par
In this way, the fundamental character of quantum laws became unquestionable. Not only because of the formal simplicity of the theory, the accuracy of its predictions, elementary systems included, but also because of an aspect of the quantum theory which is hardly ever evoked, that is,
\par
 {\textit{.. the amazing fecundity of a formalism allowing one to anticipate on experiments in a successful way, as well as to think of new realisations, one wouldn't have thought of otherwise. At the theoretical, experimental and even technological levels, it is by carrying out a reflexion within the formalism itself, that creativity often shows up in the realm of microphysics \cite{Poinat}}}~${}^{\footnotemark[1]}$.
\footnotetext[1]{It is worth noting that a striking illustration of this very aspect can be found in a recent paper of F. Wilczek \cite{Wilczek}.} 
 \par \noindent
 
\par\medskip
Now, it is also noteable that the classical physics which served as a starting point to address the
quantum phenomena, doesn't present the same fundamental character as $QM$, and should rather be derived from the latter. As is well known, this point is in itself a major issue, that of the classical/quantic frontier that will be addressed elsewhere. The difficulty, as it appears, is in the definition of such a frontier. In physics, the {\textit{criteria}} that would be necessary for such a definition are never quite universal enough. Right from the onset, and with good reasons (see \ref{SEC:3}.B.5), one thinks of the size of the physical system. However, counter-examples accumulate, and this, even beyond the size of the intermediate systems, known as the {\em{mesoscopic}} scale. In the end, the system's size, the number of particles or {\textit{quantons}} they are composed of, turn out to be quite deceptive criteria or, at the very least, not universal enough.

\par\medskip
Admittedly, there is neither definition nor any criterium that would be so universal as  to encompass the variety of boundary relations that classical and quantum physics have one to the other. In the scientific literature, this persisting and frustrating situation finds various expressions, some of them in a mild form,
\par
{\textit{This experience suggests that the problems belong to physics and that there is little chance of providing very general solutions using sophisticated mathematical machinery or philosophical considerations \cite{lindblad}}.}
\par\noindent
While others may be more extreme or provocative, as for instance \cite{mlb1},
\par
 {\textit{Existe-t-il une fronti\`ere classique/quantique?}}
 \par\noindent
 In the same line and even more recently, one may quote this passage of a book precisely devoted to the grasp of the quantum divide \cite{Gerry},
\par\medskip
{\textit{After all, it is quantum mechanics, not classical mechanics that is generally thought to be the most fundamental theory. So, perhaps, the real question is not so much about the location of the quantum/classical divide, rather than about whether there even is a divide.}}
\par\medskip\noindent
That is, from the difficulty of defining a classical/quantum frontier in a universal enough manner, one arrives at questioning the very existence of such a frontier. However, who could seriously doubt of the difference separating the classical and quantum worlds? Shouldn't this be taken, rather, as an indication that something essential is still escaping us, some {\textit{quantum enigma}} \cite{Wolfgang}? 

\par\medskip
This short paper aims at paving the way to an elucidation of the quantum enigma, positing that the clue to this longstanding interrogation is effectively not to be looked for in sophisticated mathematical machinery; but positing also that the `crazy behaviour' quantum objects do manifest, stands close to the nature of the quantum reality itself. Accordingly, an ontological questioning of this quantum nature cannot and should not be avoided \cite{Avril}.
\par
In order to carry out this project, the famous Schr\"odinger cat paradox will be considered, not with the purpose of presenting an exhaustive review of the matter, but focusing instead on the philosophical/ontological ideas which underpin the important attempts at its resolution; and to propose, thereby, an ontological questioning of the physical reality so efficiently described by Quantum Mechanics. This defines the goal of the current paper.

\par\medskip
In section \ref{SEC:2}, the paradox is quickly recalled within its historical cradle, the so-called Copenhagen interpretation of Quantum Mechanics, which the paradox was intended to test. Section \ref{SEC:3} begins with a reminder that this interpretation fails to provide the paradox with any clue, while, through the introduction of {\textit{Von Neumann chains}} more recent treatments have taken the original Schr\"odinger cat paradox to the archetypical discussion of a classical/quantum divide. The issues raised up through these attempts at solving the paradox are evaluated, and Section \ref{SEC:4} gathers conclusions that can be drawn from these, to reach the form of an ontological interrogation. A conclusion summarises the essential points of this analysis.
\par

\section{\label{SEC:2} Schr\"odinger's cat paradox}

The famous Schr\"odinger's cat paradox fits in the classical/quantum problematic, even beyond Schr\"odinger's original purposes \cite{jmll1}. In 1935, when E. Schr\"odinger proposes his case, the canonical interpretation of $QM$ - the Copenhagen interpretation - is a debated subject in itself. This interpretation, in effect, is not so well defined and makes up a rather motley ensemble. Year after year, though, it has been basically preserved until today \cite{Bachtold}, even if, sometimes, with serious provisos, as it can consistently be qualified also as a `non-interpretation', to quote A. Legget.
\par
In effect, the basic principle posited by the Copenhague interpretation is that there is no other reality than classical, and that nothing at all can be stated concerning the microphysical world. The question would even be totally irrelevant \cite{Bohr},
\par
{\textit{There is no quantum world. There is only an abstract physical description. It is wrong to think that the task of physics is to find out how Nature is. Physics is concerned with what one can say about Nature}}. 
\par
According to this view, observing or measuring would make the only reality, classical, emerge out of an objectively undetermined microphysical situation.
\par
 It is in order to probe the coherence of this vision of things, that Schr\"odinger has enunciated the school case of a cat locked in a box together with a flask of poison. Through the expediency of some evil device involving a hammer, the flask can or cannot be broken, based on whether a radioactive atom has decayed or not, after some duration of time, one hour for example.

 \par\medskip
 The paradox is ordinarily given the following expression. One considers that after one hour there is equal chances for the radioactive atom to have either decayed, and be therefore in its fundamental state, $|\Psi_1\rangle$, or to have stayed in its excited state $|\Psi_2\rangle$. The wave function of the full system, `cat plus atom', therefore reads as,
 \begin{equation}\label{chat}
 |\Psi\rangle=\frac{1}{{\sqrt{2}}}\,\bigl\lbrace \,\,|\Psi_1\rangle\!\otimes\, |dead\rangle+\,|\Psi_2\rangle\!\otimes\, |alive\rangle\,\bigr\rbrace
 \end{equation}where $a$ and $b$, complex numbers, are the process {\textit{probability amplitudes}}. This state, $|\Psi\rangle$ represents a possible state of the full system `atom plus cat'. It is a vector of the linear Hilbert space $H_a\otimes H_c$ obtained by forming the tensorial product of $H_a$ and $H_c$, the linear Hilbert spaces of atom and cat states, respectively.
 \par
 With supporting graphical illustrations, literature has often seized this wave function to represent a cat in a sort of a mixed or fuzzy  state, as if the animal was wandering about in {\textit{quantum limbos}} between life and death \cite{jmll1}, as Schr\"odinger himself was putting things,
 \par
 {\textit{In it (the wave function), the dead and alive cat are (if I dare to say so), mixed up or scrambled in equal proportions.}}
 \par\noindent
It is by observation alone, that is by opening the box \cite{Chalmers}, that the animal would escape an uncertainty which would be fixed then, once and for all, in either of the two possible states, dead or alive.

 \section{\label{SEC:3} Modern approaches}
 
 \subsection{Copenhagen Interpretation}
 
 It is interesting to begin with the Copenhague Interpretation of the paradox because, as quoted above, this interpretation furnishes the basic interpretation of Quantum Mechanics \cite{Bachtold}, never forgetting moreover that the famous paradox was inspired by the Copenhagen interpretation itself. 
 \par
 The system `atom+cat' being taken as a quantum system, the issue of knowing in which state is the cat, before the box is unlocked, is not a point. This is simply because, according to this interpretation, `{{There is no quantum world!}}'. The only one is classical.
\par\medskip
As a matter of fact, it is only by opening the box that the cat's final state is eventually {\textit{measured}}, whereas, so long as the box remains locked, the cat's state remains perfectly undetermined. What is worth emphasising here, is a singular consequence of this point of view. 
\par
The box in effect, acts as if it were some `{\textit{de-realizing}}' machine. While the entering cat is a plain real being at the beginning, after one hour of entanglement to the radioactive atom's states, the animal has become phantom-like, as if it had passed from this real world to the limbos of that {\textit{quantum non-world}}. An odd consequence indeed, rarely put forth, though worth of consideration. 

 \subsection{Von Neumann chains}
 
Now, it is important to point out that what is contained in the wave function (\ref{chat}) is not a mixture of the cat's states $|alive\rangle$ and $|dead\rangle$, but, explicitly, a mixture (a linear combination) of two states belonging to the product space $H_a\otimes H_c$, the states $|\Psi_1\rangle\!\otimes\, |dead\rangle$ and $|\Psi_2\rangle\!\otimes\, |alive\rangle$. These two states are explicitly {\textit{factorized states}}, but their sum isn't, and because of this very fact, their sum defines an entangled state : It cannot be written as a product of a state in $H_a$ by a state in $H_c$. As a result, what is to be considered as {{`mixed up'}}, if ever, is not the dead cat with the live cat, as literature often suggests, but instead the `decayed atom and dead cat' with the `non-decayed atom and live cat'. This remark, due to Ref.\cite{jmll1}, has two main virtues which are worth putting forth. 
 \par
{\textbf{ -}} First, it prevents us from considering, in an inappropriate way, any mixture of only the states `dead cat' and `alive cat', as Schr\"odinger himself did unduly (Citation, bottom of page 5).
  \par
 {\textbf{ -}} And secondly, it puts the focus on the real issue of the cat paradox, which is that of a {\em{paradigm}} as will be remarked later on.
 \par\medskip\noindent
 In order to not miss the point, it is helpful to observe that a Hilbert space, $H$, which is a linear space over the complex field, $\mathbb{C}$, cannot be used to describe the possible states of a cat, supposing that such a description could even make sense over the real field, $\mathbb{R}$~${}^{\footnotemark[2]}$.
\footnotetext[2]{To our best knowledge, the difference, complex versus real field, is never really appreciated. In Ref.\cite{jmll2}, though, an interesting discussion of the point can be found.} Though it should be intuitively obvious that the description of a cat doesn't require operators acting on any Hilbert space \cite{Gerry}, the latter statement could be provided with a more systematic foundation in a future publication. For now, suffice it to say that this statement alone would invalidate the whole affair of the Schr\"odinger's cat paradox, so long as it is formulated the standard way of equation (\ref{chat}). \par
 But, based on the fact that all of the experiment's parts are made out of {\textit{quanta}}, these sorts of objections have been historically circumvented, from upstream, in the following manner.

\par\medskip\noindent 
 A refined formulation of the paradox argues for the need that a closer look be taken at the experiment in which are involved, in addition to the cat and the atom, a Geiger counter, a hammer, a
poison bottle, etc.. This formulation will accordingly consider the possible states of the Geiger counter, its $|triggered\rangle$ and $|non-triggered\rangle$ states; then also the hammer states, $|fallen\rangle$ and $|non-fallen\rangle$, the flask's states, $|broken\rangle$ and $|non-broken\rangle$, etc. . That is, to equation (\ref{chat}), one should rather substitute something like \cite{Gerry},
 \begin{eqnarray}\label{chain}&& \nonumber
 |\Psi\rangle=\frac{1}{{\sqrt{2}}}\,\biggl\lbrace |not-decayed\rangle|not-triggered\rangle|not-fallen\rangle|alive\rangle\\ && +\,|decayed\rangle|triggered\rangle|fallen\rangle|dead\rangle\biggr\rbrace\,,
 \end{eqnarray}
 and of course, for the sake of completeness, a whole series of similar decompositions should be considered altogether. These are the so-called {\textit{Von Neumann chains}}, which come about as infinite regressions, due to the huge set of all the possible physical entanglements stretching from the atom's to the cat's states. 
 \par
 At this stage, the two following points must be retained.
 \par\noindent
 $\bullet$ However surrealistic this enumeration of all of the entangled mediations may appear (notwithstanding that these mediations can also be \emph{innumerable}), it is important to realise that for a number of physicists and philosophers, it suffers from no {{obstruction of principle}}. Quite on the contrary. As compared to equation (\ref{chat}), in effect, equation (\ref{chain}) can be considered as exhibiting a deeper degree of compliance to the physical reality of the experiment, supposing again that equation (\ref{chain}) could be written in an exhaustive, complete~form.
 \par
 What is crucial to note is that positing such an expression as equation (\ref{chain}), relies on a hidden hypothesis, which is nothing other than an unlimited application of the principle of \emph{reductionism}. In any system involving a causality, in effect, this causality must be explicable; and since ``a system is composed of parts", what cannot be explained at a certain level of details of the parts composing the system, must necessarily be explained at a more detailed level of the parts of this same system.
 \par\noindent
$\bullet$ This Von Neumann regression accounting for the full wave-function of the experiment, is now expected to yield one of either two expected results for the cat, dead or alive. The outcome having to be generated {\textit{via}} some sort of {\textit{wave-function collapse/reduction}} \cite{wavefunction}, whatever the way the collapse is generated; preferably, along the unitary evolution prescribed by the Schr\"odinger equation, in order to avoid an {\textit{ad hoc}}-reputed axiom of wave-function collapse. 
  \par\medskip
 While these two points are of a paramount importance, the latter has motivated several attempts at finding a mechanism which could account for an appropriate reduction of a Von Neumann chain such as (\ref{chain}).
  \subsubsection{The matter with unitary evolution}
Now, how do we go from a situation as complex as the one described in (2), to a dead cat or a live cat? At a theoretical level, obtaining a unique result out of the wave-function unitary evolution is the issue. The fate is that the experiment's state vector ({\textit{i.e.}}, the wavefunction) so conceived never collapses to a unique result \cite{Gerry}, and this is precisely what can be thought of as the gist of the paradox, such as posited by equation (\ref{chain}): When, where and how does the wave function reduction operate?
\par
 Here it is also that a most interesting aspect of the Schr\"odinger cat paradox shows up, which is that of a \emph{paradigm}, namely, that of the quantum/classical divide. Another interesting aspect will also be put forth in the Conclusion.
\par
Besides the very plausible incompleteness of the vector state enunciation, the difficulty has to do with a measuring process. The Von Neumann model on which one can rely when it comes to measuring a quantum system, offers an interesting starting point, but in its current state, it is unable to account for the emergence of a unique result for the measure. So long as the state vector evolution is driven by the Schr\"odinger equation, which is linear and involves a reversible evolution, the fate is that measurement doesn't yield a unique result. 
 \par
 \subsubsection{GRW theory}
 This is why, over the past decades, non-linear extensions to the Schr\"odinger equation have been considered so as to remedy this impediment. These attempts are illustrated through the GRW theory \cite{GRW} and the series of completions brought to it, including gravitation, whose unification to the quantum theories is not yet available, and the role of the observer's mind. Now, according to the experts themselves, a new universal constant would also be necessary to account for a complete theory of quantum measurement (Ref.\cite{Laloe} offers an interesting and exhaustive review of these attempts, as well as of several others). 
 \par
These tentative solutions to the wave function collapse have more recently been completed with another approach, drawn from a sound experimental {\textit{savoir faire}} \cite{Raimond, Guerlin}. Relying on a series of indirect {\textit{quantum non-demolition}} measurements, one is able to prove the convergence of the series toward the collapse of the wavefunction, such as assumed by the axioms of quantum mechanics for direct measurements \cite{Bauer1}. In doing so, there is no need for considerations other than those of repeated probe-system interactions within a pure quantum mechanical framework. Alternatively, taking things the other way round, the infinite series of indirect quantum non-demolition measurements could be viewed as building/defining an adequate measurement apparatus \cite{Bauer1}.

\par\medskip
Upstream of all of these speculations, though, it is argued elsewhere that a cogent metaphysical argument can help us evaluate how pertinent these attempts may be; or, rather, may not be \cite{wavefunction}.
\par\medskip
\subsubsection{Controlling entanglement}
 As is clear out of (\ref{chat}) and (\ref{chain}), the property of entanglement appears to be at the origin of the paradox (Conclusion, 1st paragraph). But at the level of complexity of a living animal, and well before this level is reached, it must be recalled that entanglement has escaped any possible form of theoretical control {\em{and}} conceivable detection by means of {\textit{witness operators}} \cite{Pitowsky}, or by means of {\textit{tensor-stable positive maps}}~\cite{Hermes}. 
 \par
 So that, in their practice, in order to appreciate a degree of entanglement, physicists indeed resort to guesses, and this for much smaller systems, which are genuinely quantum. Totally out of control, the infinite regression of entanglements meant by equation (\ref{chain}) appears therefore as highly speculative a statement.
\par
\subsubsection {The point of view of Decoherence Theory}
 In order to get a unique result out of a given Schr\"odinger cat experiment, it is customary to base some hopes on the {\textit{decoherence}} phenomenon \cite{Gisin2}.  As a matter of fact, in a series of experimental tests carried out on `Schr\"odinger kitties' ${}^{\footnotemark[3]}$,
\footnotetext[3]{`Kitties', because made out of only 8 intricate photons in a cavity \cite{Brune}.}linear superpositions have been proven to exist over very short lapses, $\tau_{dec}$ \cite{Brune}, provided that the quantum systems are isolated enough from their environments ${}^{\footnotemark[4]}$.
\footnotetext[4]{Durations of linear superpositions are usually extremely short, depending on the environments, as short as $10^{-30}$s, for a large molecule of radius $10^{-6}$cm, moving in the air \cite{Omnes}. Then correlations to orthogonal environment's states destroy the phase coherence of the system's states. This is called \emph{decoherence}.} \par
Even though, the problem is that decoherence transforms original linear superpositions only into so-called {\textit{improper mixtures}} \cite{mlb2}, and this means that, again, the purpose of generating a unique experimental result is missed. 
\par
More to the point, and moving upstream, a corollary of the analysis in \cite{Pitowsky} is that even though a Schr\"odinger cat state vector would stand in a superposition, that is before decoherence takes place, no witness operator would have any substantial chance to reveal it. This is because this impossibility is already effective at the much simpler level of an idealised and perfectly isolated `kitty system'. Eventually, the impossibility in question, theoretically estimated in \cite{Pitowsky}, is joined also by technical limits, such as the finite resolution power of experimental devices \cite{Gisin2}. 
\par
\subsubsection{More to the point : The Badurek experiment}
 Along the same line of approach, an {\textit{a fortiori}}-type of argument can be drawn out of the Badurek experiment \cite{mlb2}.
Before even considering the involved case of a living animal, it is instructive to remark that a simple classical field, as is the electromagnetic field, cannot get entangled to a quantum system. 
\par
The experiment consists in a spin-up neutron interferometer. In each of the branches of the interferometer, there is a constant magnetic field $B_0$, along the direction $Oz$, and a field of radio-frequency $B_1(t)$, polarized along the directtion $Ox$. The latter is oscillating at a frequency close to the Larmor frequency $\omega_0$ which makes the spins flip. Now, if the classical field $B_1(t)$ ever got entangled to the neutron's spins, the environment would keep track of the neutron's trajectories in either of the two arms of the interferometer, and as a result, interference fringes would fade away. 
 \par
 But they don't. Formally, it is possible to show \cite{mlb3} that the interaction of spins with the classical field, as described by a quasi-classical state, called {\textit{coherent state}}, preserves the coherence of the spin states $|up\rangle_1$ and $|up\rangle_2$, so that interference fringes are maintained, as experiment shows.
 \par
 Now, what is to be noted is that the coherent (or {{quasi-classical}}) state which accounts for the classical field and plays the role of the Schr\"odinger cat, is by construction close to a quantum reality, an eigenstate of the \emph{quantum annihilation operator}, satisfying the relation $ a|z\rangle=\alpha|z\rangle\,, \  \alpha \in {\mathbb{C}}$.
 This coherent state is made out of a very large number of quantum states, and it is precisely this limit that makes of the coherent state a classical field which does \emph{not} get entangled to the neutron's spin states. 
 \par
 Needless to insist at this stage that one is extremely far from the level of complexity of a real cat, and it is really surprising that such a significant case has so poorly been taken into account in the Schr\"odinger's cat paradox appraisal. 
 \par
 By the way, Equation (\ref{chain}) and related issues, appear as first instances of the so-called Everett parallel worlds whose we can glimpse here the aspect of a pure academic problem, devoid of sufficiently assured physical content ${}^{\footnotemark[5]}$.
\footnotetext[5]{Opposite points of view exist of course, taking Everett parallel worlds seriously. An analogy can be found in \cite{Tegmark}, positing Universes made out of pure mathematical relations deprived of any \emph{substratum} on which relations are normally used to hinge.}  
 \par
 In the end, the state of the cat is still unknown, and the box has to be opened in order to know it. That is, not much has really been improved on the previous case of Copenhagen interpretation. For this and other similar reasons, this famous Copenhagen interpretation, with appropriate adaptations, has been able to survive until today \cite{Bachtold}.  
  
  \subsubsection{Everett parallel worlds}
  Everett's interpretation of Quantum Mechanics seems to have the favor of a certain number of {\textit{cosmologists}} whose considerations bear on the Universe wave-function ${}^{\footnotemark[6]}$.
\footnotetext[6]{As pointed out earlier, an unlimited validity of the reductionism principle is assumed here again.} Admittedly, though, this approach meets quite serious, yet unsolved difficulties \cite{Laloe}. 
  \par
  Everett's interpretation takes seriously the existence of Von Neumann chains, and posits that they furnish a direct and sound representation of the physical reality \cite{Laloe}. As can be found in a number of specialised articles, the entanglement mechanism is assumed to extend, step by step, to the whole environment \cite{lindblad}; up to, and included, the experimenters themselves, taken to be as entangled as is the cat of the famous Schr\"odinger case.
  \par
  To sum up, Von Neumann chains ramify and suffuse infinitely. These chains lead to a multitude of different Universes, that one is bound to think of as being {\em{parallel}} to our Universe. Effectively, it is no longer just the cat and the atom which form a quantum system, but also the laboratory where they are studied, and the experimenters, and the city, and the country, and the land, {\textit{etc}}.
\par
Therefore, there is no need for any wave-function collapse mechanism that would fix the experimental result in one of two final possibilities, because all intermediate possibilities get their own Universe. They all get realised and the Von Neumann chain is not interrupted. In the end, there are as many  Universes, parallel the ones to the others, as ramifications undergone by the wave-function.
  \par
A variant of this interpretation posits that the state vector ramifications are no longer realised in a `multi-world' production, but are realised in the observers own minds, while observers own states themselves get entangled to the other states of the Von Neumann chain. Within one and the same Everett interpretation of Quantum Mechanics, these two variants may be regarded as somewhat reminiscent of the long {\textit{epistemological-vs-ontological}} issue concerning the wave-function status \cite{wavefunction}.
\par
Now, what does all this amount to, in consideration of the cat's affairs? Clearly, one cannot help agreeing that Everett's interpretation is most appealing to the animal, which can freely escape to a better world in which radioactive atoms do not decay, while hammers fall down on flasks without breaking them, not to speak of entanglement to physicists' tortuous minds. With Everett, one can therefore hope that the cat of Mr. Schr\"odinger is now purring in one of these better worlds.
\par
 However, if one opens the box, either a dead or a live cat is found, and this, whatever the number of times the experiment is run. And in this way one can see dramatically enough that Everett's parallel worlds do not seem to offer a safe escape from the terrible alternative.
\par
One may object that within this interpretation, in another world corresponding to another ramification of the state vector (the chain), the radioactive atom did not decay, and accordingly, neither did the cat. \par
But, which cat? After one hour, opening the box, a cat is found, the one that entered the box. Has the other state vector ramification generated the exact copy of Mr. Schr\"odinger's cat in an alternate Universe? Were it alone, a famous theorem of quantum physics would object to that ${}^{\footnotemark[7]}$.
\footnotetext[7]{That is the \emph{non-cloning theorem},  \cite{noncloning}}  
Are the state vector ramifications accompanied with new worlds spontaneous creation?..  in our immediate neighborhood?.. without us perceiving the slightest echo of these gigantic creations?..  The fate is that the branching of the World is unobservable~\cite{Baggott}.
\par
One could say that indeed, the multiverse explanation always comes up against the same difficulty, that of positing the ramifications of the world, which have never been observed, for lack of being able to account, by unitary evolution, for the wave function's reduction, which, it, on the contrary, is constantly verified at the macroscopic level.

\subsubsection{Another variant of Everett parallel worlds: Griffiths consistent histories}
Instead of a many Universe  interpretation,  M. Gell-Mann has advocated an alternative interpretation of Everett's work in terms of `many alternative \emph{histories}', unfolding in one and the same Universe. The histories under consideration were defined by R. B. Griffiths as quantum mechanical consistent sequences of events, unfolding along a given temporal sequence \cite{Griffiths}. This approach has been supported by a number of major experts \cite{Baggott}, and has even been claimed, by R.B. Griffith, to be `Copenhagen done right' ${}^{\footnotemark[8]}$.
\footnotetext[8]{This is because consistent histories lend themselves to an explicit translation of Bohr complementary principle in terms of probabilities \cite{Baggott}.} But considering families of histories, an example of which is (\ref{chain}) itself, no physical law is known to select a would-be `correct family of histories'.  A consistent history like (\ref{chain}) (or any of its {\em{truncature} } at a given step) is recorded in this formalism, but not really questioned regarding either its selection or its pertinenece. In the end, the consistent histories approach wouldn't shed any new light on the cat paradox, and like the other attempts, this approach tries to circumvent what definitely appears as the real theoretical problem posed by the Schr\"odinger cat paradox, the wave function's reduction driven by a unitary evolution.

\section{\label{SEC:4} Discussion}

Since it was launched by E. Schr\"odinger in the 1930s, the cat paradox has been the matter
of an important number of publications and speculations, and this until nowadays \cite{Gerry, Gisin2}. Now, having reviewed the main attempts at its resolution, one is allowed to think that a resolution of the cat’s school case is not achieved yet.

In the cat paradox, and all of the subsequent attempts at its resolution, the property of entanglement, which E. Schr\"odinger's himself considered as the most salient feature of quantum physics, plays a central role and is taken for granted from the onset ${}^{\footnotemark[9]}$
\footnotetext[9]{Entanglement should rather be viewed a consequence of the superposition
principle, the only mystery of quantum physics, as pointed out rightly by R.P. Feynman \cite{Gerry}.}
\par\medskip

- At an epistemological level now, it is important to notice that entanglement
displays a capacity proper to a quantum entity, to bear the potentiality of a property in
common with other quantum entities. The notion of potentiality {\em{must}} be used here in order to recall
that a quantum object does not possess its properties in an \emph{actual} manner : The result of
a measure doesn't pre-exist the measuring procedure, or, in other words, is not revealed by
the measure, as in the classical case. Rather, it is induced through, and along the measure
itself \cite{wavefunction}, and this quantic peculiarity applies to all of the quantum entities, entangled or not.

It may be opportune to scrutinise this point a bit further. As is well known, and has been the matter of a great number of experiments carried out since the mid-seventies, two elementary particles are able to bear together the potentiality of one and the same measurable degree of freedom, such as spin and/or polarisation.

But even in this situation, most appropriate to the matter of entanglement, care
must be taken not to be fooled by words and to keep tight control of what we talk about.

Elementary particles are known as plain quantum entities \cite{PEs}, and they do not have real
and actual proper existences, aside from the experimental protocols which define them as
such \cite{Gerryp.53}. Strictly \cite{Haag}. This is even more blatant with particles in entangled states. In a
much deeper understanding therefore, when the proper reality of elementary particles is considered, the latter
must be referred to that particular state of the quantum field associated to them \cite{dEspagnat}; and it
is it, this particular state of the field, which will account for all of the subsequent measurements which,
in a second step only, will be described in terms of elementary particles in order to stick to the
overall intuitive scheme that physicists have in mind \cite{PEs}.

Of course, one is now able to prepare `kitties' and `squids' (superconducting quantum interference devices involving the mesoscopic scale of some $10^6$ electrons) which, though bigger
and bigger, still manifest quantum behaviours such as linear superposition, entanglement and
interferences \cite{Gisin1}. But aside from being still extremely far from a cat, the entangled elements composing these kitties and squids do not enjoy a full actualised reality, possessing their properties in an actual manner. Unlike a real cat .. even the Schrodinger’s cat.

Things unravel as soon as they are taken from a perspective along
which it is the nature of the particles (deeply: the nature of the quantum fields associated
to them) which displays a property, entanglement, to which linear algebra offers a simple
and accurate mathematical formulation, so long as a few degrees of freedom are involved \cite{mlb1} ${}^{\footnotemark[10]}$.
\footnotetext[10]{As bigger numbers of degrees of freedom get involved, possibly infinite, the formalism of relativistic quantum field theories may be used, in particular that of $C^\star$-opeator algebras. Quantum fields entanglement, then, reveals to be an even more significant property than that of smaller systems, ordinarily described within the standard formalism of Quantum Mechanics \cite{Halfvorson}.}

From a historical point of view, of course, the notion of entanglement came out of the
Quantum Mechanics formalism in the first place, and was experimentally confirmed afterwards
\cite{aspect}. As evoked in the Introduction, this is precisely an amazing instance of the fecundity of
the formalism, an interpretation of which will be worth proposing elsewhere. Now, if the
$QM$ formalism has proven to be a place for anticipation/creativity, this quality
shouldn't be taken to the extreme point of positing that the formalism would precede reality,
in a typical platonic way. It is this attitude, however, that is implicit to the way in which
the Schr\"odinger's cat paradox is posed.

In effect, as stated in \cite{Gisin2} `.. nothing in the QM formalism prevents one from applying
it to macroscopical objects.', and this is what is done with the cat in the famous paradox.
But it is done recklessly. Contrarily to quantum entities, in effect, a cat possesses the plain actuality
of its determinations, and in order to describe a cat, there is no need to substitute operators
and state vectors for numbers and functions.

In support of this, we have seen that even a classical magnetic field doesn't get entangled to neutron's spins,
and, {\textit{a fortiori}}, it is the same in the case of a cat with respect to the states of a radioactive
atom. The only outcome for this system, considering the atom state's superposition, is that
after a given duration of time, there is a certain probability for the animal to be either dead
or alive. But this is in no way indicative that the cat would have passed from a supposedly
entangled state (would this be possible) to a factorised one. Final state statistics are the
same in either cases \cite{mlb}, a generic situation in Quantum Mechanics, known under the name
of non-uniqueness of the preparation; this cannot stand for a proof of entanglement, not even evoking the
impossibility of any theoretical control and reliable detection of entanglement, beyond a
certain level of complexity.
\par\medskip
- At a theoretical level, it is usually believed that a resolution of generic Schr\"odinger's cat
situations would require that a theory of quantum measurement be sufficiently achieved in
order to end up with some definite and unique result out of Von Neumann chains, like expression (2) ${}^{\footnotemark[11]}$.
\footnotetext[11]{A different point of view could be exposed elsewhere on the bases of further developments.}
\par

Besides, decades of tremendous developments, relying on all sorts of extensions to the
original linear problem, and including gravitation, have not produced convincing and universal
solutions. It is then not surprising that, in a recurrent fashion since the early days of quantum physics, in
addition to gravity whose unification to quantum physics is not yet available, some authors
go back again to the observer's conscience in order to find a solution~\cite{Penrose}.
\par

In view of these `over-speculative' considerations, one may think instead that the resolution
of the quantum enigma is not to be looked for in these directions, and that it may
stand closer to the nature of the quantum object itself \cite{Grandou}.

\section{\label{SEC:5} Conclusion}

Finally, the school case of Schr\"odinger, though designed as `perfectly burlesque' by E. Schr\"odinger himself, entails two points which are fundamental and stand at the core of the quantum enigma. 
\par\medskip
$\bullet$ A first point is that by enunciating such a case, one relies on a hidden implicit assumption, that of an unlimited principle of {\textit{reductionism}}. Clearly, equation (\ref{chain}) is based on this assumption. This is also stated explicitly in the citation given above, that `.. nothing in the Quantum Mechanics formalism prevents one from applying it to macroscopical objects.'  That new levels of reality may have emerged between the atom and the cat, in the line of `More is different', is simply not envisaged \cite{moreisdifferent}.
\par
$\bullet$  A second point has to do with three facts the paradox confronts us with. As coherence holds for the radio active atom, the atom {\textit{is}} in a superposition of two states, decayed and non-decayed (f1). As coherence is over, and some state vector reduction has taken place, the atom {\textit{is}} decayed (f2). As the atom has decayed, the cat {\textit{is}} dead (f3). The three facts are all three \emph{real} facts, and it is one of the merits of the Schr\"odinger school case to force us to understand that {\em{to be}} in (f1) is as real as {\em{to be}} in the second (f2) and third one (f3). In other words, the Schr\"odinger's cat paradox works as a paradigm of what is at stake with the quantum reality: How is it that something real, something `that is', is in the manner of a superposition? The cat is not the problem, for we know for sure that it is either alive or dead. 
\par
The atom is the problem, and the formalism of Quantum Mechanics conveys this problem in the most revealing manner. It is in the very nature of a quantum reality to possess the {\em{real}} property of {\em{existing}} in many orthogonal actualities.
\par\medskip
Having said that, the Schr\"odinger's cat paradox appealed to all from the beginning because it linked a quantum reality to a classical one, pretending that since the former exerted a causality on the state of the latter, they were entangled as every other entangled quantum systems. The cat was therefore summoned to inherit a quantum nature from the mere fact of its dependence on a quantum behaviour. 

\par
Once placed in the box, it was no more a common cat, which must be either dead or alive, but it entered a new mode of being where cats are dead and alive at the same time, in the same manner as quantum entities \emph{are} in a state of superposition.
\par
This of course cannot stand. But this common sense observation was obfuscated when the paradox was formulated, because of the condition that the box would remain sealed for a certain duration of time.
\par
Being in a state of linear superposition is a strange mode of being compared to the actual mode of being of cats, and we must not escape the question this strange mode of being raises. It is not everyday that physics gets the privilege to confront us with such a puzzle, which is metaphysical by essence.

\begin{acknowledgments}
It is a pleasure to thank Fr. Dennis Klein, O.P., who helped us improving some formulations.
\end{acknowledgments}


\begin{thebibliography}{}
%
%


\bibitem{Gisin1}
N. Gisin, `L'impensable Hasard', Odile Jacob, 2012.


\bibitem{Gleason} A. M. Gleason,(1957). "Measures on the closed subspaces of a Hilbert space". Indiana University Mathematics Journal. 6 (4): 885–893. doi:10.1512/iumj.1957.6.56050. MR 0096113.

\bibitem{KSKC}
S. Kochen and E. Specker,  “The Problem of Hidden Variables in Quantum Mechanics”, Journal of Mathematics and Mechanics, 17 (1967), 59.  J. Conway and S. Kochen, `Free will theorem', Notices of the American Society, Vol.56, No.2 (2009). 

\bibitem{Poinat}
S. Poinat, PhD thesis, University Nice Sophia-Antipolis, 08/12/2009.


\bibitem{Wilczek} F. Wilczek,  `Quanta of the Third Kind', Inferences 6(3) (2021); https://inference-review.com/article/quanta-of-the-third-kind




\bibitem{lindblad}
G. Lindblad, `Foundations of quantum mechanics?', Phys. Scr.84 (2011) 018501.


\bibitem{mlb1}
M. Le Bellac, `Quantum World', World Scientific, 2013.




\bibitem{Gerry}
C.C. Gerry and K. M. Bruno, `The Quantum Divide',  Oxford, 2013, p.100.

\bibitem{Wolfgang}
W. Smith, `The Quantum Enigma', Sherwood Sugden and Company,1995.

\bibitem{Avril} A. Styrman, Journal of Physics,  Conf. Ser. 1466 012001(2020).

\bibitem{jmll1}
`Le chat de Schr\"odinger', J-M. Levy-Leblond, Sciences \& Avenir, Ao\^ut 2006.

\bibitem{Bachtold}
M. B\"achtold, `L'interpretation de la M\'ecanique Quantique, une approche pragmatiste',
Vision des Sciences, Hermann, 2008.

\bibitem{Bohr}
N. Bohr, `Letter to Aage Peterson', cited in Ref.\cite{Poinat}.

\bibitem{Chalmers}
M. Chalmers, `State of mind', New Scientist, May 2014.

\bibitem{jmll2}
F. Balibar, A. Laverne,
J-M. L\'evy-Leblond and D. Mouhanna, Quantique : Elements, 2007,  https://cel.archives-ouvertes.fr/cel-00136189/document.

\bibitem{wavefunction}
D-M. Cabaret, T. Grandou and E. Perrier,  `Status of the wave-function of Quantum Mechanics', Foundations of Physics.
{|\bf{DOI}} : 10.1007/s10701-022-00574-w; arXiv:2103.05504v1 [physics.hist-ph].

\bibitem{GRW}
G.C. Ghirardi, A. Rimini and T. Weber, Phys. Rev. D 34, (1986), 470.

\bibitem{Laloe}
F. Laloe, `Comprenons-nous vraiment la M\'ecanique Quantique?', p.275, CNRS Editions, 2011, p.275.


\bibitem{Raimond}
J.M. Raimond, M. Brune and S. Haroche, Phys. Rev. Lett. 79 (1997) 1964.

\bibitem{Guerlin}
C. Guerlin et al., Nature 448 (2007) 889.


\bibitem{Bauer1}
M. Bauer and D. Bernard, `Convergence of repeated quantum non-demolition measurements and wave function collapse', Phys. Rev. A 84, (2011) 044103, arXiv: 1106.4953v2 [math-ph].

\bibitem{Pitowsky} I. Pitowsky,
`Macroscopic objects in quantum mechanics: A combinatorial approach', Phys. Rev. A 70 (2004) 022103.

\bibitem{Hermes}
A. M\"uller-Hermes, D. Reeb and M.M. Wolf, `Positivity of linear maps under tensor powers', J. Math. Phys. 57 (2016) 015202.

\bibitem{Gisin2}
`Le chat de Scrodinger devient r\'eel', La Recherche Juillet-Aout 2015, p.53.

\bibitem{Brune}
J.M. Raimond, M. Brune and S. Haroche, Phys. Rev. Lett. 79 (1997), 1964. 

\bibitem{mlb2}
M. Le Bellac, Physique Quantique, Tome I, 3rd edition, CNRS Editions, 2013. English translation of 1st edition, `Quantum Physics', Cambridge University Press, 2007.


\bibitem{Omnes} R. Omn\`es, The interpretation of quantum mechanics, Princeton University Press, 1994.

\bibitem{mlb3}
M. Le Bellac, Physique Quantique, Tome II, p.767, 3rd edition, CNRS Editions, 2013.



\bibitem{Tegmark} M. Tegmark,
La Recherche, No 489 Juillet-Aout 2014, p.24. `L'essence du monde est math\'ematique'.

 

\bibitem{noncloning}
`The non-cloning theorem', M. Le Bellac, Physique Quantique, Tome II, 3rd edition, CNRS Editions 2013.

\bibitem{Baggott} J. Baggott,`beyond measure', Oxford University Press, 2004, p.266.

\bibitem{Griffiths} D. Griffiths, Introduction to Quantum Mechanics, 2nd Edition, Pearson Prentice Hall, 2004.

\bibitem{PEs}
D-M. Cabaret, T. Grandou, G.-M. Grange and E. Perrier, `Elementary particles: What are they? Substances, elements and primary matter',   Foundations of Science (2022), https://doi.org/10.1007/s10699-021-09826-w; arXiv:2103.05522v1 [physics.hist-ph]. 
\bibitem{Gerryp.53}
See Ref.\cite{Gerry} on page 53.

\bibitem{Haag}
R. Haag, in `Operator Algebras and Quantum Statistical Mechanics', O. Bratteli and D.W. Robinson Eds., Springer, 1987.


\bibitem{dEspagnat}
B. d'Espagnat, `Physique contemporaine et intelligibilt\'e du monde', in PhiloScience, $N^o1$, p.5, Universit\'e Interdisciplinaire de Paris, Hiver-Printemps 2004-2005.


\bibitem{Halfvorson}
`Entanglement and Open Systems in Algebraic Quantum Field Theory', R. Clifton and H. Halvorson, 	Stud.Hist.Philos.Mod.Phys. 32 (2001) 1, arXiv:quant-ph/0001107v1.

\bibitem{aspect}
A. Aspect, P. Grangier and G. Roger, Experimental realization of Einstein-Podolsky-Rosen gedanken experiment: A new violation of Bell's inequalities, Phys. Rev. Lett. {\bf{49}} (1982), 91. 

\bibitem{mlb} 
Cf. Ref.\cite{noncloning}, p. 391.


\bibitem{Penrose}
R. Penrose, `Shadows of the mind', Oxford Press, 1994.


\bibitem{Grandou}
T. Grandou, in `Les Sciences face \`a la Cr\'eation', ICES Editions, 2014.

\bibitem{moreisdifferent}
P. W. Anderson. `More is different', Science, New Series, Vol. 177, No. 4047 (1972), 393.

















































































\end{thebibliography}
\end{document}